\documentclass[aps,pra,preprint,showpacs,groupedaddress,superscriptaddress,longbibliography]{revtex4-1}

\usepackage[english]{babel}
\usepackage{float}
\usepackage{graphicx}
\usepackage[caption=false]{subfig}
\usepackage{amsmath}
\usepackage{amsfonts}
\usepackage{amssymb}
\usepackage{mathtools}
\usepackage{amsthm,enumitem}

\begin{document}

\title{Spectral Properties of Confining Superexponential Potentials}

 \author{Peter Schmelcher}
  \email{Peter.Schmelcher@physnet.uni-hamburg.de}
 \affiliation{Zentrum f\"ur Optische Quantentechnologien, Universit\"at Hamburg, Luruper Chaussee 149, 22761 Hamburg, Germany}
 \affiliation{The Hamburg Centre for Ultrafast Imaging, Universit\"at Hamburg, Luruper Chaussee 149, 22761 Hamburg, Germany}

\date{\today}

\begin{abstract}
We explore the spectral properties and behaviour of confining superexponential potentials.
Several prototypes of these highly nonlinear potentials are analyzed in terms of the eigenvalues
and eigenstates of the underlying stationary Schr\"odinger equation up to several hundreds of
excited states. A generalization of the superexponential self-interacting oscillator shows a 
scaling behaviour of the spacing of the eigenvalues which turns into an alternating behaviour
for the power law modified oscillator. Superexponential potentials with an oscillating power
show a very rich spectral structure with varying amplitudes and wave vectors.
In the parity symmetric case doublets of near degenerate energy eigenvalues emerge
in the spectrum. The corresponding eigenstates are strongly localized in the outer wells
of the potential and occur as even-odd pairs which are interspersed into the spectrum of
delocalized states. We provide an outlook on future perspectives including the possibility to
use these features for applications in e.g. cold atom physics.
\end{abstract}

\maketitle

\section{Motivation and Introduction} 
\label{sec:introduction}

\noindent
Confining particles to a given spatial region opens the doorway for their controlled
preparation and processing. It allows for an efficient detection as 
well as measurement of the properties and interactions of ensembles of particles.
This holds for a large variety of systems, including atoms and molecules as well as
larger clusters of particles or even, molecular motors, cells and bacteria.
Consequently a plethora of different possibilities
emerge to investigate the structure and dynamics of the constituents of matter, including
their response to external fields. In different fields of physics and chemistry confining
or trapping particles paves the way for e.g. high resolution spectroscopy of cold and
ultracold molecules \cite{Smith,Krems}, optical tweezer based manipulation of soft materials 
\cite{Ho} and the exploration of ultracold quantum matter and Bose-Einstein condensates
\cite{Pethick}.

\noindent
The importance of the geometry of the external trapping potential becomes particularly 
impactful in the case of ultracold quantum matter \cite{Metcalf,Grimm,Pethick}. Laser and evaporative cooling
in a trapping environment leads in the ultracold regime to the formation of degenerate atomic quantum gases.
The geometry of the external trapping potential created by inhomogeneous static electric and magnetic 
and/or laser fields can nowadays be shaped almost arbitrarily ranging from box-like traps,
harmonic oscillator or double well confinement to periodic optical lattices. 
These traps, in combination with the control of the interatomic interactions 
\cite{Chin}, imprint and probe different properties of the correlated few- and many-particle systems
under investigation. A box-like trap allows to probe the physics, such as sound dispersion and
soliton formation \cite{Pethick,Frantzeskakis} or the probing of the equation of state
\cite{Moritz}, of a homogeneous systems. Harmonic and double well confinement allow to probe
the collective modes \cite{Pethick} and interaction-induced tunneling mechanisms \cite{Schmelcher1} as well
as tunneling and nonlinear self-trapping in a bosonic Josephson junction \cite{Oberthaler}.
On the other hand optical lattices represent a unique platform for exploring weakly
to strongly correlated many-body systems \cite{Bloch} with a plethora of quantum phases appearing
with increasing complexity, the paradigm being the superfluid Mott-insulator quantum phase transition
\cite{Greiner}. This way a close bridge is established between condensed matter systems and ultracold
quantum gases.

\noindent
In a different direction of research very recently there has been some first explorations of so-called
superexponential systems \cite{Schmelcher2,Schmelcher3,Schmelcher4,Schmelcher5}.
These are model systems where the underlying exponential potential exhibits a spatial 
dependence for both the base and the exponent. A prototype potential is ${\cal{V}} = |q_1|^{q_2}$ where 
$q_1$ and $q_2$ are the coordinates of particles. In spite of its simple appearance
such a two-body potential leads to a rich geometrical structure. Changing the exponent
degree of freedom $q_2$ from $-\infty$ to $+\infty$ one encounters a power law confining channel for $q_2>0$
with a continuously changing value for the power that transits via two saddle points to a
region of asymptotically free motion \cite{Schmelcher2}. The resulting scattering dynamics 
reflects this transition with increasing energy. In the many-body case 
multiple backscattering and recollision events due to an intermittent behaviour in the
saddle point regime have been observed \cite{Schmelcher3}. For the case of only a single
degree of freedom $q$ the very peculiar properties of the so-called self-interacting
superexponential oscillator (SSO) with the potential ${\cal{V}} = |q|^q$ have been analyzed
\cite{Schmelcher4}. In contrast to the (an-)harmonic oscillator the SSO shows an exponentially
varying nonlinearity. Its potential exhibits a transition point with a hierarchy
of singularities of logarithmic and power law character leaving their fingerprints
in the agglomeration of its phase space curves. As a consequence the period of the
SSO undergoes a crossover from decreasing linear to a nonlinearly increasing
behaviour with increasing energy. The quantum SSO shows some remarkable spectral and eigenstate
properties according to the quantum signatures of this classical transition \cite{Schmelcher5}.
The ground state undergoes a metamorphosis of decentering, squeezing and the emergence of a tail.
For energies below the transition point a scaling behaviour in the spectrum was detected 
for large amplitudes of the SSO.

\noindent
While the above-discussed features of superexponential systems demonstrate 
novel structures and dynamics, the question arises what interesting spectral features
do exist for superexponential quantum systems beyond the quantum SSO. To address this question
we focus in the present work on confining superexponential potentials in a single spatial
dimension. We will analyze the spectrum and eigenstate properties for a wide range of excitations
for several geometrically appealing potential landscapes. We derive scaling properties of the
eigenvalues of these highly nonlinear potentials and analyze as well as characterize the 'dynamics' of the eigenvalue
spacings with increasing degree of excitation. In particular we demonstrate that confining
oscillatory superexponential potentials exhibit interspersed localized and delocalized states.
The localized states occur in doublets which represent energetically near degenerate pairs of eigenstates
that live in one of the outer wells of the superexponential potential. The two partners
of these doublets possess opposite parity. Even if the parity symmetry is broken and consequently
the near degeneracy is lifted by inducing a corresponding phase shift of the potential 
the localization is still maintained and the corresponding eigenstates are exclusively left or
right localized. Based on the above spectral features we provide a discussion of future perspectives
and possible applications of confining superexponential potentials.

\noindent
This work is organized as follows. In section \ref{sec:qsso} we explore the spectral properties
of the SSO with varying amplitude. Section \ref{sec:rsso} contains a spectral analysis of a symmetrized
version of the SSO. The power law enhanced SSO is addressed in section \ref{sec:psso}. 
Oscillating power potentials are explored and analyzed in the following section \ref{sec:opp}
where we start with the sublinear case for specific phases. The eigenvalue spectrum for arbitrary phases
is discussed and we demonstrate the symmetry breaking process. Finally we show spectral results for the
quadratic and briefly for the quartic case. Section \ref{sec:concl} contains our conclusions and an outlook
including future perspectives.

\section{Quantum SSO with varying amplitude} 
\label{sec:qsso}

\begin{figure}
\includegraphics[width=8cm,height=6cm]{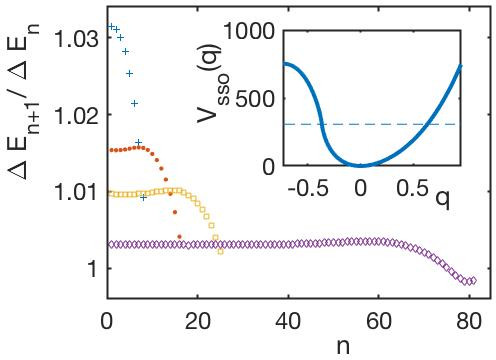}
\caption{The scaled spacing $\frac{\Delta E_{n+1}}{\Delta E_n}$ of the eigenenergies of the SSO with varying
excitation label $n$ for different values of the amplitude $\alpha$: From top to bottom
we have $\alpha=10^3, 4 \cdot 10^3, 10^4, 10^5$. Inset: the spatially and energetically shifted SSO potential
well as given in the text for $\alpha = 10^3$; the dashed line indicates the energy of the transition point.}
\label{Fig1}
\end{figure}

\noindent
As a first step in our investigation of the spectral properties and structure of confining superexponential
potentials let us briefly address the quantum SSO with varying amplitude $\alpha$. The quantum SSO has been
investigated in quite some detail in ref.\cite{Schmelcher5} but with a focus on very large amplitudes. Therefore,
we provide here, as a piece of complementary information, the spectral behaviour also for intermediate and small
amplitudes. The Hamiltonian reads

\begin{equation}
{\cal{H}} = \frac{p^2}{2m} + \alpha |q|^q
\end{equation}

\noindent
and we assume, without loss of generality,
in the following $\hbar=m=1$ while varying the amplitude $\alpha$. The potential ${\cal{V}}= \alpha |q|^q$
is highly nonlinear and asymmetric and its potential well possesses no reflection symmetry around its minimum
(see inset of Figure \ref{Fig1} for the shifted potential well as discussed below). The minimum and maximum are located
at $q_{min}= \left( \frac{1}{e} \right)$ with ${\cal{V}}(q_{min})=e^{-\frac{1}{e}}$ and at
$q_{max}= \left(- \frac{1}{e} \right)$ with ${\cal{V}}(q_{max})=e^{\frac{1}{e}}$, respectively. Half way between
the minimum and the maximum a transition point occurs at $q=0$ (see ref. \cite{Schmelcher5} for details)
where all derivatives of ${\cal{V}}$ are singular. Shifting the potential energy and the position
in coordinate space of the minimum to zero yields the shifted SSO potential 

\begin{equation}
{\cal{V}}_{sh}(q) = \alpha \left( |q+e^{-1}|^{q+e^{-1}} - e^{-e^{-1}} \right)
\end{equation}

\noindent
We focus in this work on the eigenstates and eigenenergies being solutions to the corresponding
stationary Schr\"odinger equation (SEQ) ${\cal{H}} \psi_n = E_n \psi_n$. A few comments concerning our numerical
approach to solve the SEQ are in order.
To obtain the eigenvalues and eigenstates of the SEQ for several hundred
excited eigenstates we use an eighth order finite difference discretization scheme of space \cite{Groenenboom}.
Determining the eigenvalues and eigenvectors then corresponds to the diagonalization of the resulting
Hamiltonian band matrix.

\noindent
We concentrate on the spectral properties below the transition point energy (see inset of Figure \ref{Fig1})
which exhibits an intriguing scaling behaviour as we shall see below. Figure \ref{Fig1} shows the scaled
spacing $\Delta R_n=\frac{\Delta E_{n+1}}{\Delta E_n}$, with $\Delta E_n = E_n - E_{n-1}$ being the difference of the
eigenenergies of the $n-$th and the $n-1$st eigenstates, as a function of the degree of excitation
for varying amplitude $\alpha$ of the SSO.
First of all one observes that the values of the scaled spacing are always close to one, varying only by
a few percent around one. For the case $\alpha = 10^3$ with increasing degree of excitation the scaled spacing
monotonically decreases ranging for low excitations in the vicinity of $1.03$ and for states close
to the transition energy around $1.01$. For larger values of $\alpha$ a plateau emerges with approximately constant
values for $\Delta R$ with $\Delta R = 1.015, 1.01, 1.003$ for $\alpha = 4 \cdot 10^3, 10^4, 10^5$, respectively.
This clearly indicates that there is a scaling property of the eigenvalue spectrum in the well of the SSO,
and the scaling factor decreases towards the value one with increasing amplitude $\alpha$ of the SSO.
The constancy of this scaling over a broad range of the spectrum is a remarkable property of the SSO.

\noindent
A natural modification of the SSO potential is the skewed superexponential oscillator with the potential
${\cal{V}}_{sk} (q) = \alpha |\beta q|^{\gamma q}$. We remark that this skewed oscillator shows similar
spectral properties as compared to the original SSO, and we therefore refrain from discussing it separately.

\section{The right-symmetrized SSO} 
\label{sec:rsso}

\begin{figure}
\includegraphics[width=8cm,height=6cm]{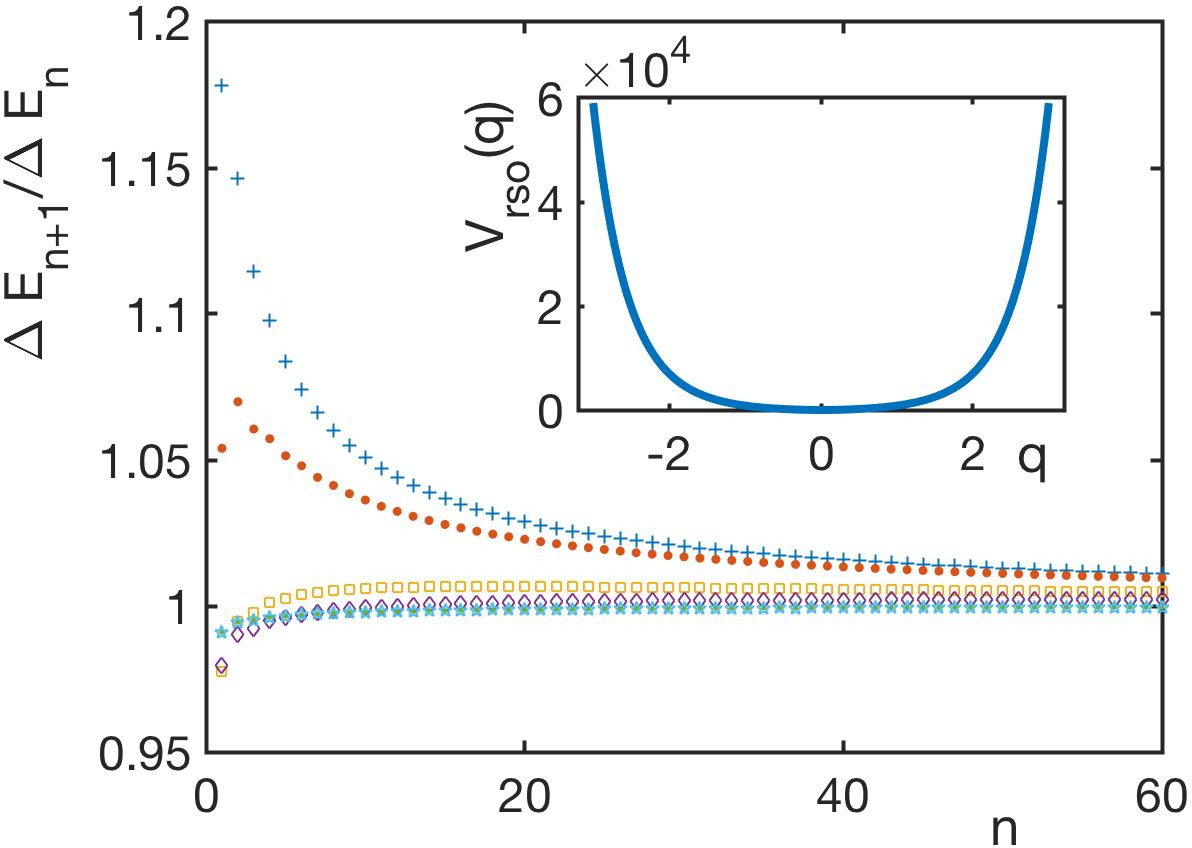}
\caption{The scaled spacing $\Delta R_n=\frac{\Delta E_{n+1}}{\Delta E_n}$ of the eigenenergies of the right-symmetrized SSO with varying
excitation label $n$ for different values of the amplitude $\alpha$: From top to bottom
we have $\alpha=1, 10, 10^3, 10^4, 10^6$. Inset: the potential ${\cal{V}}_{rso}$ of the right-symmetrized SSO
as given in the text for $\alpha = 10^3$.}
\label{Fig2}
\end{figure}

\noindent
A modification of the SSO, which introduces new properties into the spectrum, is obtained if we simply mirror the 
right half of the SSO potential well to the left and obtain in this way a reflection symmetry potential
well. This is the so-called right-symmetrized SSO. Shifting it to zero energy and to the minimum position
at the origin one obtains

\begin{equation}
{\cal{V}}_{rso} (q) = \alpha \left( \left(q_m + |q| \right)^{\left(q_m + |q| \right)} - q_m^{q_m} \right)
\end{equation}

\noindent
where $q_m=e^{-1}$. The potential well of the right-symmetrized SSO is shown in the inset of Figure \ref{Fig2}
for $\alpha =10^3$. 

\noindent
Figure \ref{Fig2} shows the scaled spacing $\Delta R_n$ of the eigenenergies of the right-symmetrized
SSO as a function of the degree of excitation $n$ for varying amplitude $\alpha$ of the SSO. It shows
a smooth approach with (in magnitude)
monotonically decreasing slopes for the different values of $\alpha$ towards the asymptotics.
For small values of $\alpha$, namely $\alpha = 1,10$, $\Delta R_n$ is larger than the value one and
decreases monotonically with increasing $n$ towards the values one. Opposite to this, for $\alpha = 10^6$
$\Delta R_n$ is always less than one and increases monotonically towards one. For the intermediate cases
$\alpha = 10^3, 10^4$, $\Delta R_n$ increases with increasing $n$ from values smaller than one, then passes
the threshold value of one and increases further. Subsequently (barely visible in Figure \ref{Fig2}) it reaches
a maximum (e.g. $n=18$ for $\alpha = 10^3$) and decreases towards the value one with further increasing $n$.
This clearly demonstrates that the right-symmetrized SSO and the SSO spectrum behave qualitatively different.
This provides us with the promise of an even richer spectral structure when moving on to other superexponential trapping
potentials. Indeed we will see this conjecture being verified below for other natural generalizations of the
SSO.

\section{The power law SSO} 
\label{sec:psso}

\begin{figure}
\includegraphics[width=8cm,height=6cm]{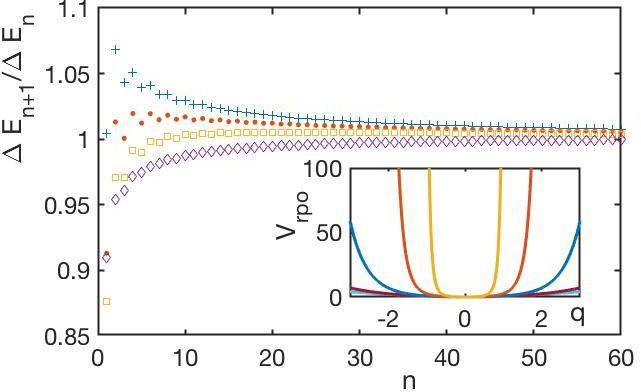}
\caption{The scaled spacing $\Delta R_n=\frac{\Delta E_{n+1}}{\Delta E_n}$ of the eigenenergies of the power
law SSO with varying excitation label $n$ for $\beta=0.5$ and different values of the amplitude $\alpha$: From top to bottom
$\alpha=1, 10, 10^2, 10^4$. Inset: the potential ${\cal{V}}_{rpo}$ of the RPO 
as given in the text for $\alpha = 1, \beta=0.25,0.5,1.0,2.0,4.0$ from bottom to top.}
\label{Fig3}
\end{figure}

\begin{figure}
\includegraphics[width=8cm,height=6cm]{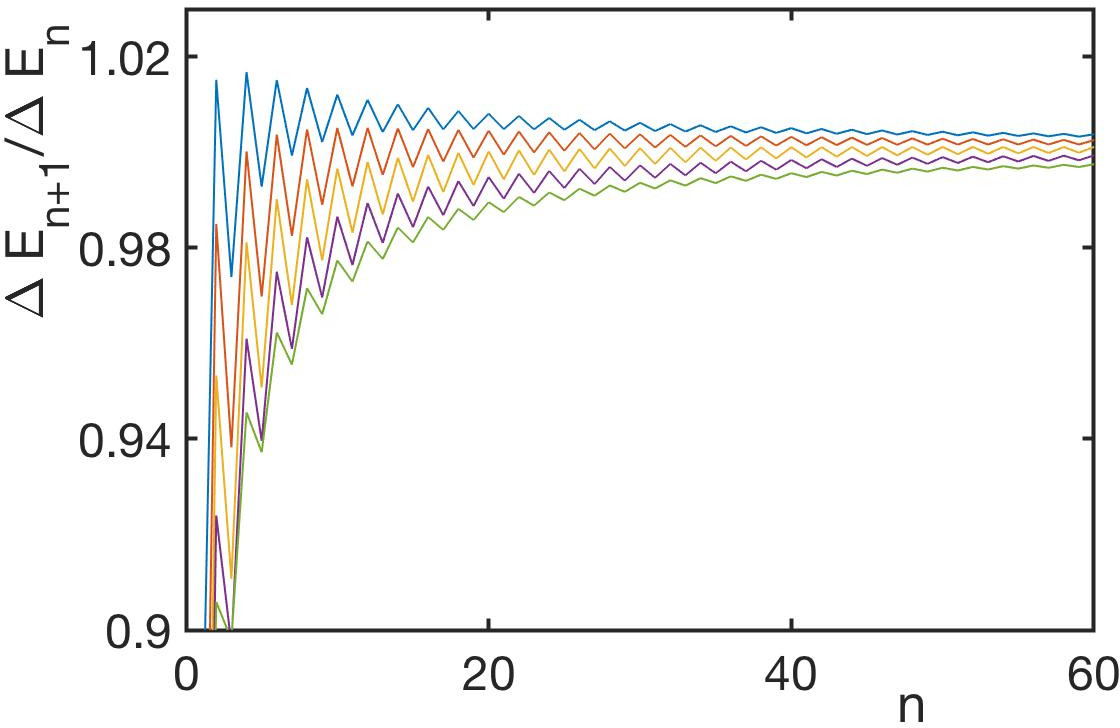}
\caption{The scaled spacing $\Delta R_n=\frac{\Delta E_{n+1}}{\Delta E_n}$ of the eigenenergies of the RPO with varying
excitation label $n$ for $\beta=0.25$ and different values of the amplitude $\alpha$: From top to bottom
$\alpha=1, 10, 10^2, 10^3, 10^4$. To guide the eye the data points have been interconnected with straight
lines.}
\label{Fig4}
\end{figure}

\noindent
Let us now consider the right-symmetrized SSO with an extra power to possibly tune the spectral properties. The
corresponding potential of this power law SSO reads as follows

\begin{equation}
{\cal{V}}_{rpo} (q) = \alpha \left( \left(q_m + |q| \right)^{\left(\left(q_m + |q| \right)^{\beta} \right)} - 
q_m^{(q_m^{\beta})} \right)
\end{equation}

\noindent
where now $q_m = e^{-\frac{1}{\beta}}$ and we have two free parameters: the amplitude $\alpha$ and the
exponent $\beta$ in the exponent. The inset of Figure \ref{Fig3} provides the potential ${\cal{V}}_{rpo}(q)$
for $\alpha=1$ with varying values of the parameter $\beta$. The increasingly steep confinement with
an increasing value of $\beta$ is clearly visible. Figure \ref{Fig3} shows the scaled spacing $\Delta R_n$
of the eigenenergies of the power law SSO with varying excitation label $n$ for $\beta = 0.5$ and for
different values of the amplitude $\alpha$. The spectral behaviour is similar to the case of the right-symmetrized
SSO discussed in the previous section, but now additionally we observe for low excitations that the monotonic
behaviour of $\Delta R_n$ with increasing $n$ is violated: we encounter an increase versus decrease 
for the scaled spacing for consecutive values of $n$. This is much more pronounced for smaller values
of the parameter $\beta$. Figure \ref{Fig4} addresses the case $\beta=0.25$ where in an alternating
manner throughout the spectrum an increase is followed by a decrease of the scaled spacing $\Delta R_n$.
In certain parts of the spectrum these oscillations even refer to values of $\Delta R_n$ above and 
below the value one.

\section{The oscillating power potential} 
\label{sec:opp}

\noindent
Let us now allow for an oscillating periodic function in the exponent while we keep as a basis the magnitude
of the coordinate $q$. We therefore introduce the oscillating power potentials (OPP) according to the
following 

\begin{eqnarray}
{\cal{V}}_c (\alpha,\beta,k;q)= |q|^{\alpha + \beta \cos(kq)}\\
{\cal{V}}_s (\alpha,\beta,k;q)= |q|^{\alpha + \beta \sin(kq)}
\end{eqnarray}

\begin{figure}
\includegraphics[width=7cm,height=6cm]{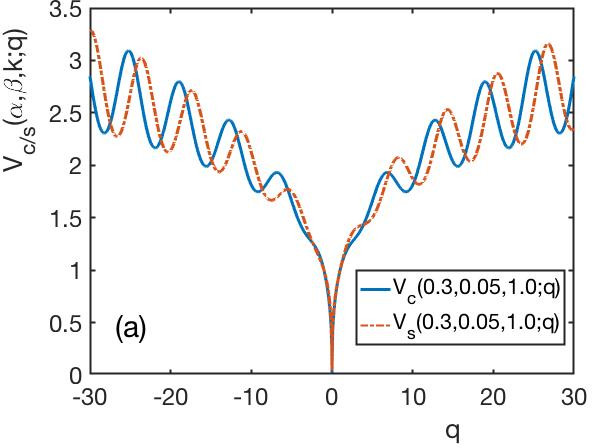}
\includegraphics[width=7cm,height=6cm]{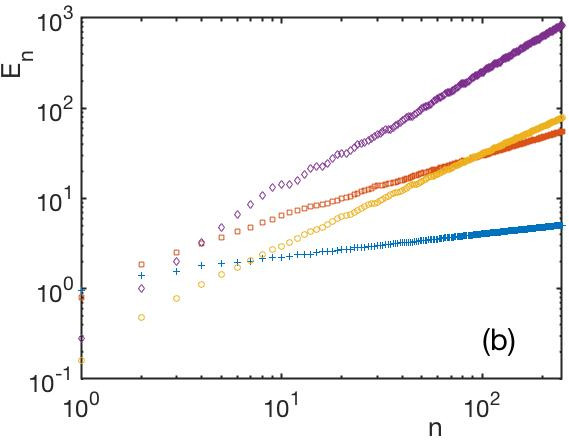}\\
\includegraphics[width=7cm,height=6cm]{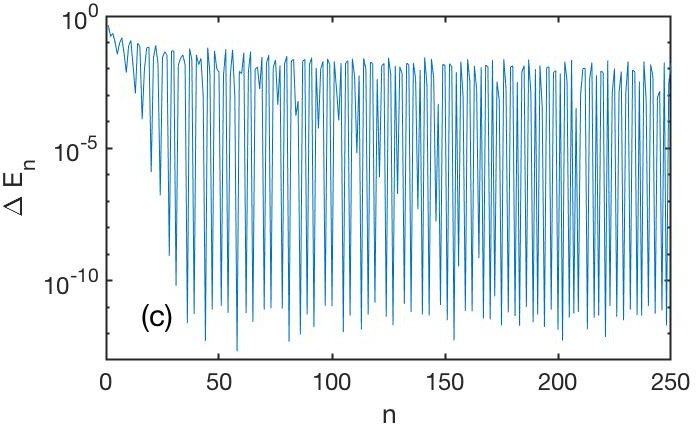}
\includegraphics[width=7cm,height=6cm]{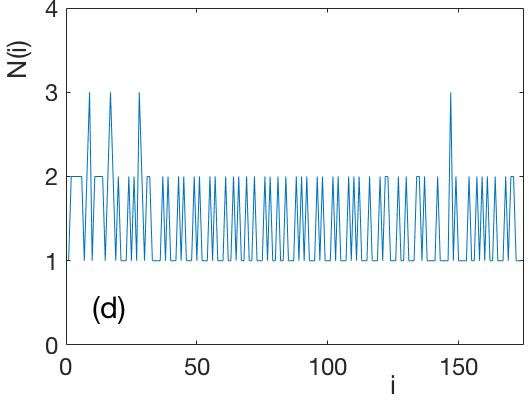}
\caption{(a) The potentials ${\cal{V}}_c(\alpha,\beta,k;q)$ and correspondingly
${\cal{V}}_s(\alpha,\beta,k;q)$ (see text) for $\alpha=0.3,\beta=0.05, k=1$.
(b) Log-log representation of the energy eigenvalues $E_n$ versus the excitation
label $n$ for $\alpha=0.3,\beta=0.05,k=1,\gamma=1$ (crosses),
$\alpha=1,\beta=0.05,k=1,\gamma=1$ (squares),
$\alpha=2,\beta=0.1,k=3,\gamma=0.05$ (circles),
$\alpha=4,\beta=1,k=2,\gamma=0.1$ (diamonds). 
(c) The spacing $\Delta E_n$ of the eigenenergies as a function of the excitation label $n$
for the potential ${\cal{V}}_c(\alpha,\beta,k;q)$ with $\alpha=0.3, \beta=0.05, k=1$
on a semilogarithmic scale.
(d) The turning point characteristics $N(i)$ of the energy spacing dynamics (definition see text) shown in (c).}
\label{Fig5}
\end{figure}

\noindent
These potentials are obviously very much different from a standard superposition of a harmonic confinement
with a periodic lattice such as $V \propto q^2 + \cos(q)$. Indeed, the OPP possess an oscillating power 
with amplitude $\beta$ and wavevector $k$ superimposed on a constant power $\alpha$. Figure \ref{Fig5}(a)
shows ${\cal{V}}_c$ for the parameter values $\alpha=0.3,\beta=0.05, k=1$. Apart from the existence of 
a central well with a cusp there is a series of decentered wells which become increasingly deeper with
increasing distance from the outer well. The energies of the minima of those wells increase according to
the constant power $|q|^{0.3}$ while moving away from the origin.  These are a few relevant,
but certainly not the only, properties which distinguish ${\cal{V}}_c$ from the above-mentioned
case $V$: the OPP is a highly nonlinear potential via its oscillatory exponential appearance.
Obviously, ${\cal{V}}_c$ is reflection symmetric around
the origin, whereas ${\cal{V}}_s$ is asymmetric due to the odd character of the sine function in the
exponent (see Figure \ref{Fig5}(a)).

\subsection{The sublinear OPP cosine case}
\label{sec:opp1}

\noindent
Let us now focus on the spectral analysis of the OPP according to the solutions of the underlying SEQ.
First we will discuss the eigenvalue spectrum and consequently the most important properties of the
eigenstates.
Figure \ref{Fig5}(b) shows the energy eigenvalues $E_n$ versus the excitation label $n$
for the $250$ energetically lowest eigenstates in a double logarithmic representation for the
case $\alpha=0.3,\beta=0.05, k=1$ and for $\gamma = 1$ where $\gamma$ is an additional
prefactor (amplitude) of the overall potential. As expected,
according to the constant power in the exponent, the mean behaviour of the spectrum
follows a power law according to $E \propto n^{0.26}$.
Figure \ref{Fig5}(c) shows the spacing $\Delta E_n$ with varying degree of excitation $n$
on a semilogarithmic scale. First of all one observes an overall decrease of the spacings
with increasing $n$. Beyond this, the eyecatching feature is the highly oscillatory character
over many orders of magnitude of this spacing 'dynamics'. Starting from the ground state
energy and increasing the value of $n$ we observe a series of turning points in the
spacing dynamics (see Figure \ref{Fig5}(c))
which belong to increasingly lower values of the spacings, i.e. a series of near degeneracies
of two neighboring energy levels occur in the spectrum. In between these near degeneracies
the spacing returns to values of the order of $10^{-2}$.  For $n>20$ those near degenerate levels
involve spacings smaller than $10^{-10}$ and cannot be completely resolved within our
numerical approach. For $n>60$ a second series of increasingly narrower near degeneracies emerges
with further increasing degree of excitation $n$. As a result the 'density' of near degeneracies
in the spectrum is enhanced. 

\noindent
To gain more insights into the oscillatory dynamics of the energy spacing with increasing degree
of excitation we employ a turning point analysis. Let us define the turning point characteristics
$N(i)$ as the number of level spacings between two turning points of the level spacing 
dynamics. The index $i$ represents just the increasing natural numbers counting these level spacings
between turning points. A value of one means that two turning points follow upon each other
without the presence of a second spacing whereas a value of two indicates the presence of a second
spacing before the next turning point occurs. In Figure \ref{Fig5}(d) we observe that the values one and
two dominate $N(i)$: the dynamics shows trains of alternating values one and two occasionally 
interrupted by intervals showing two consecutive values one. Very rarely high peaks with the value three occur,
i.e. the spacing increases/decreases consecutively among four eigenenergies following upon each other.
This energy spacing analysis provides a comprehensive view on the level dynamics and we will use
it in the following to characterize the spectral properties of the different superexponential potentials.

\begin{figure}
\includegraphics[width=8cm,height=6cm]{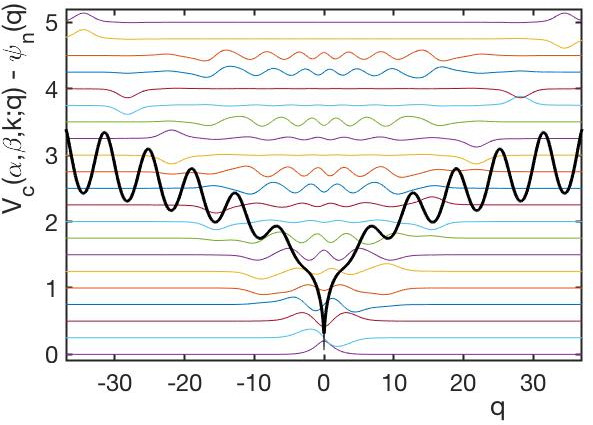}
\caption{The potential ${\cal{V}}_c(\alpha,\beta,k;q)$ for $\alpha=0.3, \beta=0.05, k=1$
together with the energetically lowest $21$ eigenstates (shown are the non-normalized probability amplitudes).
Note that the vertical arrangement of the eigenstates is not associated with the detailed energetical position
of these eigenstates.}
\label{Fig6}
\end{figure}

\noindent
Let us now inspect the eigenstates of the sublinear OPP cosine case. This will provide us with an understanding
of the origin of the near degeneracies observed above. Figure \ref{Fig6} shows the potential
${\cal{V}}_c(\alpha,\beta,k;q)$ for $\alpha=0.3, \beta=0.05, k=1$ together with the energetically
lowest $21$ eigenstates. Starting from the ground state the energetically lowest four states are
localized in the central well of the OPP. Thereafter, with further increasing degree of excitation,
more and more outer wells are covered with increasing energy. A closer inspection reveals that there is
two classes of states. The first class is the expected class of delocalized eigenstates covering a certain
range of the OPP. Interspersed into these delocalized states we encounter localized states that carry
a significant probability amplitude only in some outer wells. They occur in pairs with even and odd
parity. In Figure \ref{Fig6} these are, e.g. the pairs of the $(9,10)-$th states, or, showing an 
even stronger localization, the $(13,14)-$th, $(16,17)-$ or $(20,21)-$st eigenstates.
The localization happens
in (or above) a certain outer well and becomes increasingly more pronounced with increasing degree of
excitation of the considered eigenstate. In essence, these localized states appear to consist, to
some degree of approximation, of two single humped probability amplitudes localized at the positions
of certain outer wells whose depths increases with increasing distance from the central well and 
therefore the degree of localization becomes increasingly more pronounced. 

\subsection{The sublinear OPP sine case}
\label{sec:opp2}

\noindent
Let us now explore the spectral properties of the OPP ${\cal{V}}_s(\alpha,\beta,k;q)$. It has no
inversion symmetry w.r.t. the origin and therefore the eigenstates are not parity symmetric. 
In other words, the inversion symmetry ${\cal{V}}_c(\alpha,\beta,k;q)={\cal{V}}_c(\alpha,\beta,k;-q)$
is maximally broken for ${\cal{V}}_s$, see also Figure \ref{Fig5}(a) for the case
$\alpha=0.3, \beta=0.05, k=1$. At equal distance from the origin
the left and right outer wells are (approximately) shifted by a phase of $\frac{\pi}{2}$ i.e. if one encounters a 
maximum on the left branch of the OPP the reflection symmetric position to the right represents a minimum
and vice versa. 

\noindent
The smooth mean part of the behaviour of the eigenenergies with increasing degree of excitation is for
the OPP ${\cal{V}}_s$ the same as the one for ${\cal{V}}_c$, i.e. it obeys a power law $E \propto n^{0.26}$.
Major differences however are revealed when focusing on the fluctuations or the level spacing dynamics.

\begin{figure}
\includegraphics[width=7cm,height=6cm]{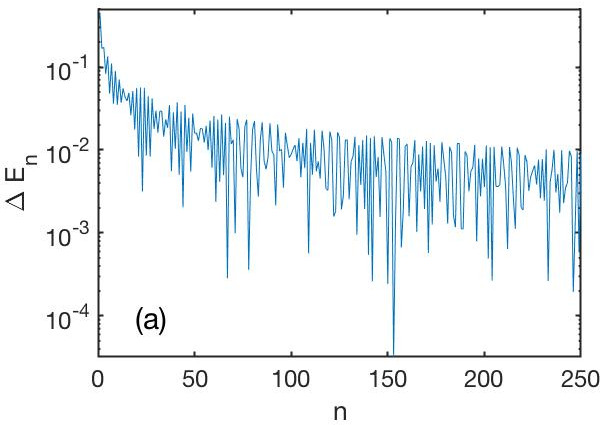}
\includegraphics[width=7cm,height=6cm]{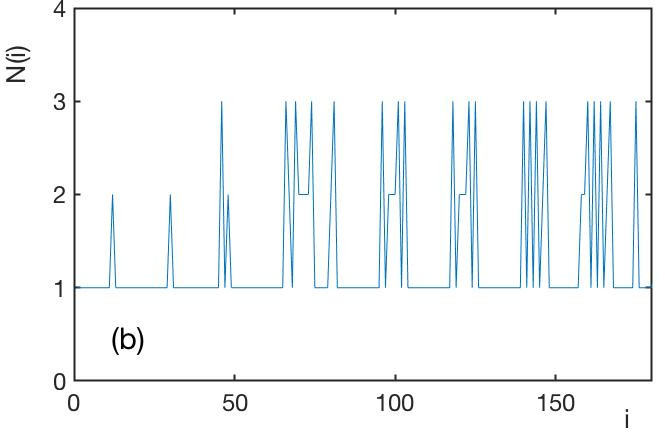}
\caption{(a) The spacing $\Delta E_n$ of the eigenenergies as a function of the excitation label $n$
for the potential ${\cal{V}}_s(\alpha,\beta,k;q)$ for $\alpha=0.3, \beta=0.05, k=1$
on a semilogarithmic scale and (b) the turning point characteristics $N(i)$ of the energy spacing
dynamics (definition see text) shown in (a).}
\label{Fig7}
\end{figure}

\noindent
Figure \ref{Fig7}(a) shows the spacing $\Delta E_n$ of the eigenenergies as a function of the excitation label $n$
for the potential ${\cal{V}}_s(\alpha,\beta,k;q)$ for $\alpha=0.3, \beta=0.05, k=1$
on a semilogarithmic scale. Obviously a very rich energy spacing dynamics can be observed.
Apart from an expected overall decay of the spacing we observe a beating behaviour and, within these
beats i.e. on the shortest scale, alternating spacings of increasing and decreasing values
between nearest neighbor spacings or next to nearest neighbor spacings.
The beats dissolve respectively overlap for larger values of the excitation $n$.
While the spectrum of the OPP ${\cal{V}}_c(\alpha,\beta,k;q)$ (see subsection \ref{sec:opp1})
contains a large number or series of near degeneracies this is now not the case 
for the spectrum of the OPP ${\cal{V}}_c(\alpha,\beta,k;q)$: there is only a very limited range of values for the 
spacings, typically $10^{-4}<\Delta E_n < 10^{-1}$, i.e. the near degeneracies of ${\cal{V}}_c$ are 
lifted due to the broken parity symmetry. Figure \ref{Fig7}(b) shows the turning point
characteristics $N(i)$ of the energy spacing dynamics. Compared to Figure \ref{Fig5}(d), which
shows $N(i)$ for the OPP ${\cal{V}}_c$, we encounter now many subsequent events with a single
spacing. The reader should be reminded that the value one for $N(i)$ corresponds to a single spacing
between turning points of the energy level dynamics, i.e. an alternating increase and decrease of the
spacing. In between these extended intervals of value one, $N(i)$ shows finite series of peaks with 
values two and three arranged in a repeating manner.

\begin{figure}
\includegraphics[width=8cm,height=6cm]{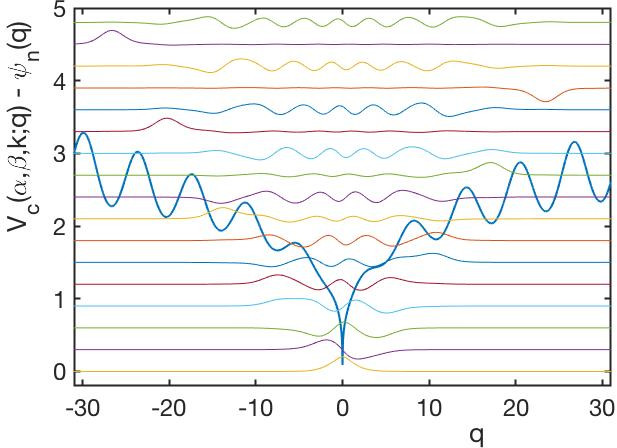}
\caption{The potential ${\cal{V}}_s(\alpha,\beta,k;q)$ for $\alpha=0.3, \beta=0.05, k=1$
together with the energetically lowest $17$ eigenstates (shown are the non-normalized probability amplitudes).}
\label{Fig8}
\end{figure}

\noindent
Figure \ref{Fig8} shows the potential ${\cal{V}}_s(\alpha,\beta,k;q)$ for $\alpha=0.3, \beta=0.05, k=1$
together with the probability amplitude of the energetically lowest $17$ eigenstates. As for the case of
${\cal{V}}_c$ (see subsection \ref{sec:opp1}) there are localized states interspersed between delocalized
states. Now, however, due to the missing parity symmetry of the eigenstates there is either right or left
localized states. In Figure \ref{Fig8} these are the $10,12,14,16-$th eigenstate which can be, according
to their localization, assigned to the corresponding outer potential wells on either the left or the
right half of the OPP.

\subsection{The sublinear OPP case with arbitrary phase}
\label{sec:opp3}

\begin{figure}
\includegraphics[width=8cm,height=6cm]{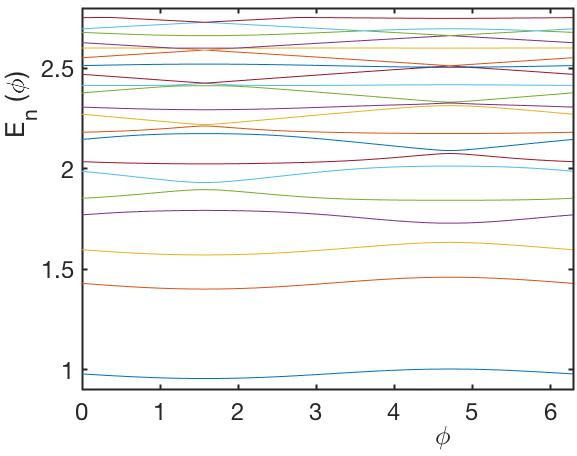}
\caption{The eigenvalue spectrum as a function of the phase $\phi$ (in arc measure) for the potential 
${\cal{V}}_p=|q|^{0.3+0.05 \sin (q+\phi)}$ for the energetically lowest 21 eigenstates.}
\label{Fig9}
\end{figure}

\noindent
Having explored the spectral properties of ${\cal{V}}_c$ and ${\cal{V}}_s$ including the presence and
lifting of near degeneracies as well as the localization and delocalization of the underlying eigenstates,
our focus in this brief subsection is to mediate between the two cases. To this end we investigate the spectrum
of the potential

\begin{equation}
{\cal{V}}_p (\alpha,\beta,k,\phi;q) = |q|^{\alpha + \beta \sin(q+\phi)}
\end{equation}

\noindent
which coincides with ${\cal{V}}_c$ and ${\cal{V}}_s$ for $\phi=\frac{\pi}{2}$ and $\phi =0$, respectively.
We again focus on the parameter values $\alpha = 0.3, \beta=0.05, k=1$ and vary $\phi$ continuously from $0$ to $2\pi$.
This way we expect to see the crossover between the spectral properties of the limiting cases considered
above. Figure \ref{Fig9} shows the eigenvalue spectrum as a function of the phase $\phi$ for  
${\cal{V}}_p$ for the energetically lowest 21 eigenstates. It can be seen that near degeneracies and
avoided crossings, which become increasingly narrow with increasing degree of excitation,
occur at $\phi = \frac{\pi}{2},\frac{3 \pi}{2}$ and are lifted once moving off from these configurations.
At $\phi =0, \pi, 2 \pi$ these close encounters of the energy levels are not present throughout the
spectrum. The doublet formation and splitting can therefore be controlled by varying the phase $\phi$.
This is somehow reminescent of the splitting of the single particle eigenstates in a double well
\cite{Schmelcher1,Oberthaler} which determines the frequency of the Rabi oscillations when considering
the dynamics following a wave packet in a single well. In our superexponential setup the situation
is different in the following sense. With increasing degree of excitation or increasing energy, 
the eigenstates of the doublets refer, in terms of their localization, to different spatially well-separated
individual outer wells attached to the overall confining potential well. As a consequence the 
resulting quantum dynamics of a wave packet, being left or right localized due to a superposition
of the two states of the doublet, would provide a quantum state transfer between remote
pairs of individual wells attached to the confining walls of the overall potential
(see also remarks in the conclusions). It should also be noted that the shape of those
individual wells and consequently also of the corresponding localized wave packets differ 
significantly from the one of the individual wells commonly encountered in optical lattices and correspondingly
from their Wannier states.

\subsection{The linear OPP cosine case}
\label{sec:opp4}

\begin{figure}
\includegraphics[width=7cm,height=6cm]{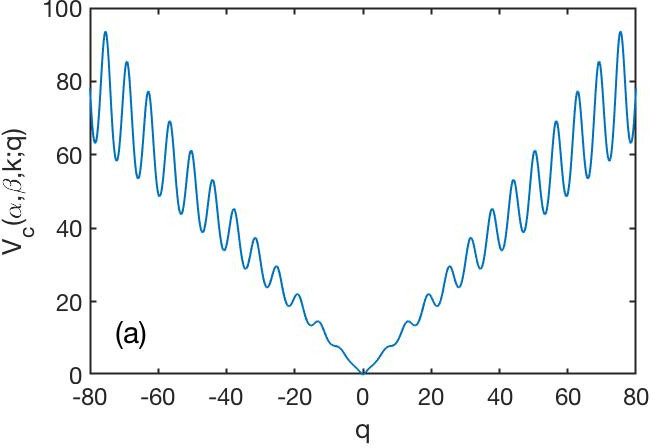}
\includegraphics[width=7cm,height=6cm]{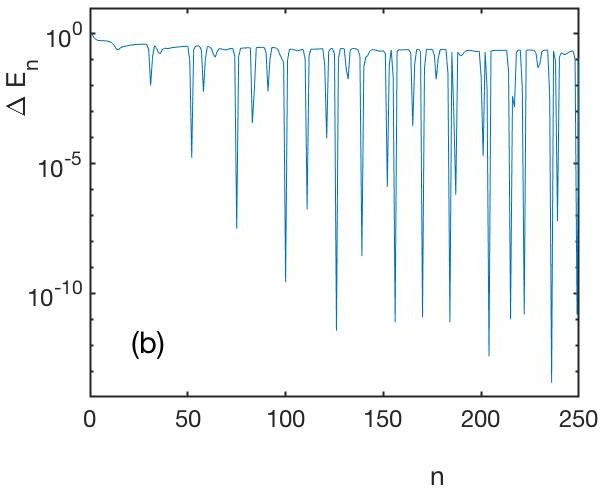}\\
\includegraphics[width=7cm,height=6cm]{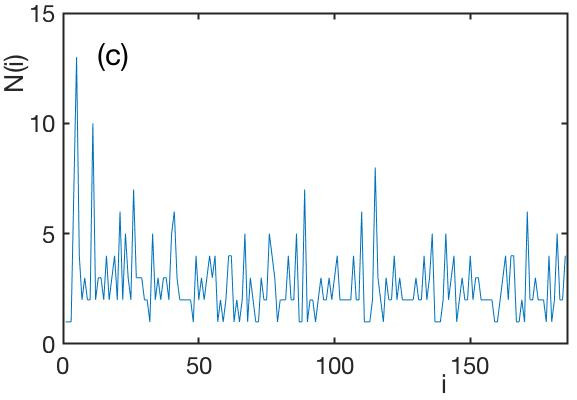}
\includegraphics[width=7cm,height=6cm]{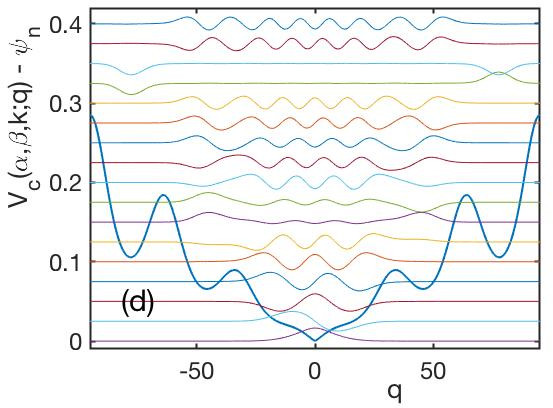}
\caption{(a) The potential ${\cal{V}}_c(\alpha,\beta,k;q)$ (see text) for $\alpha=1.0,\beta=0.05, k=1$.
(b) Semi-logarithmic representation of the spacing of the energy eigenvalues $E_n$ versus the excitation
label $n$ for the potential shown in (a). (c) The corresponding
turning point characteristics $N(i)$ of the energy spacing dynamics (definition see text) shown in (b).
(d) The potential $\gamma {\cal{V}}_c(\alpha,\beta,k;q)$ for $\alpha=1.0, \beta=0.09, k=0.2, \gamma=0.002$
together with the energetically lowest $17$ eigenstates (shown are the non-normalized probability amplitudes).}
\label{Fig10}
\end{figure}

\noindent
In this subsection we explore the case $\alpha = 1$ i.e. the case of a linear behaviour due to the
constant part of the exponent combined with an oscillating cosine in the exponent again with 
an amplitude $\beta = 0.05$. The corresponding potential ${\cal{V}}_c(\alpha,\beta,k;q)$ is shown
for $k=1$ in Figure \ref{Fig10}(a). Figure \ref{Fig5}(b) shows the mean power law dependence
of the eigenenergies with increasing excitation label for which $E \propto n^{0.72}$ holds.
Figure \ref{Fig10}(b) provides a semi-logarithmic representation of the corresponding spacing $\Delta E_n$
of the energy eigenvalues versus the excitation label $n$. First of all we observe a very 'slow'
overall decrease of the spacing with increasing degree of excitation $n$. More important, however, is
the occurence of near degeneracies which are again interspersed into the spectrum of spacings whose
predominant range of values is $10^{-2}<\Delta E_n<1$.

\noindent
Figure \ref{Fig10}(c) shows the corresponding
turning point characteristics $N(i)$ of the energy spacing dynamics. Opposite to the above considered
cases of superexponential potentials a direct alternation in the spectrum corresponding to a value of one,
is now encountered very rarely. Instead, there is now a range of values up to $13$ which is covered in the
considered window of the eigenenergy spectrum. Overall $N(i)$ looks rather irregular while still certain
patterns or bursts of spacing sequences are recognizable. Figure \ref{Fig10}(d) presents the
potential $\gamma {\cal{V}}_c(\alpha,\beta,k;q)$ for $\alpha=1.0, \beta=0.09, k=0.2, \gamma=0.002$
and its energetically lowest $17$ eigenstates. Delocalized states and states localized in outer wells,
being parity symmetric or antisymmetric, can be identified as for the above-discussed cases of the OPP.

\subsection{The quadratic and quartic OPP cosine case}
\label{sec:opp5}

\begin{figure}
\includegraphics[width=7cm,height=6cm]{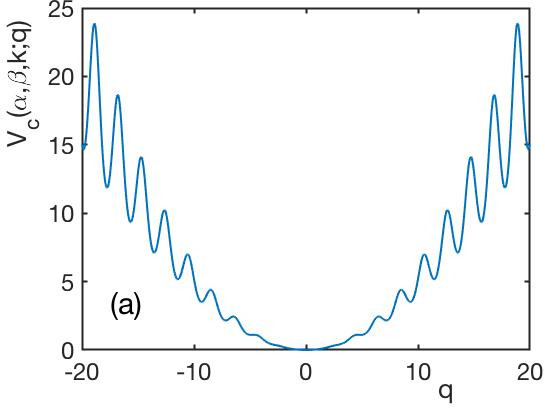}
\includegraphics[width=7cm,height=6cm]{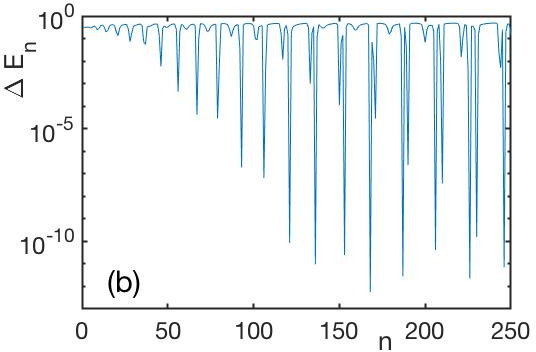}\\
\includegraphics[width=7cm,height=6cm]{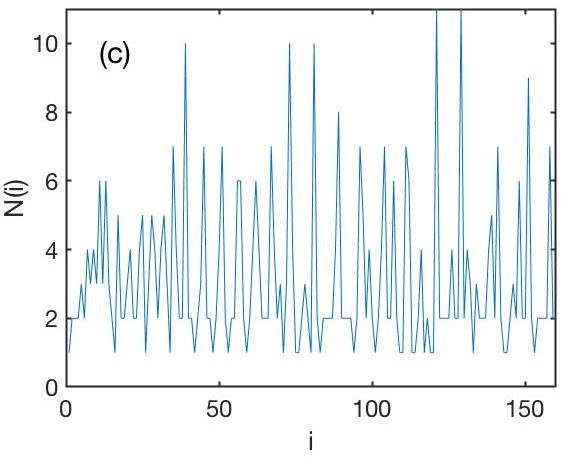}
\caption{(a) The potential $\gamma {\cal{V}}_c(\alpha,\beta,k;q)$ for $\alpha=2.0, \beta=0.1, k=3.0, \gamma=0.05$.
(b) Semi-logarithmic representation of the spacing of the energy eigenvalues $E_n$ versus the excitation
label $n$ for the potential shown in (a). (c) The corresponding turning point characteristics $N(i)$ 
of the energy spacing dynamics (definition see text) shown in (b).}
\label{Fig11}
\end{figure}

\noindent
Let us briefly study the spectral properties for the quadratic and quartic case, i.e.
for $\alpha =2$ respectively $\alpha = 4$. In Figure \ref{Fig11}(a) the potential
$\gamma \cdot {\cal{V}}_c(\alpha,\beta,k;q)$ for $\alpha=2.0, \beta=0.1, k=3.0, \gamma=0.05$
is illustrated. The deepening of the potential wells attached to the (average) potential walls,
which represent a harmonic oscillator, are clearly visible. We therefore expect also for this reflection
symmetric case the appearance of near degenerate doublets in the energy spectrum.
Indeed inspecting the dynamics of the spacing of the energy eigenvalues shown in Figure \ref{Fig11}(b)
one observes with increasing degree of excitation $n$ the emergence of near degeneracies. 
More specifically, there is two such series of near degeneracies, the first one starting at
approximately $n =50$ and second one arising close to $n=120$ (the onset of a third series
is visible too for $n > 200$).
The mean behaviour of $E_n$, as expected, follows a power law $E_n \propto n$ (see Figure \ref{Fig5}(b))
and therefore the spacing fluctuates around a constant value (see Figure \ref{Fig11}(b)).
The corresponding turning point characteristics $N(i)$ is illustrated in Figure \ref{Fig11}(c).
Many equally spaced events with values $N(i)>2$ can be observed for higher excitations, whereas
for lower excitations more rapid oscillations also for $N(i)>2$ are observed. $N(i)$ bears quite
some similarities but also major differences to the linear case discussed in subsection \ref{sec:opp4},
whereas it is overall distinctly different from the sublinear case discussed in subsections
\ref{sec:opp},\ref{sec:opp1},\ref{sec:opp2},\ref{sec:opp3}.

\begin{figure}
\includegraphics[width=8cm,height=6cm]{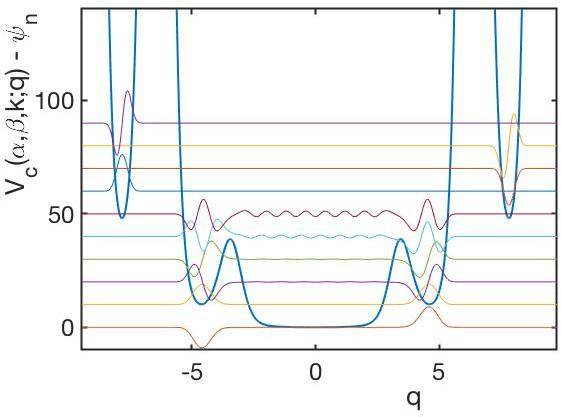}
\caption{The potential $\gamma {\cal{V}}_c(\alpha,\beta,k;q)$ for $\alpha=4.0, \beta=1.0, k=2.0, \gamma = 0.1$
together with the localized states corresponding to the $10-11th, 15-16th, 20-21st, 34-35th, 43-44th$
pairs of eigenstates (shown are the non-normalized probability amplitudes).}
\label{Fig12}
\end{figure}

\noindent
Figure \ref{Fig12} shows, as a final case, the potential $\gamma {\cal{V}}_c(\alpha,\beta,k;q)$ for 
$\alpha=4.0, \beta=1.0, k=2.0, \gamma = 0.1$ where the first few 'side-pockets' attached to the steep walls of the
overall $q^4$-confinement are clearly visible. Also shown are the localized eigenstates in this 
potential which correspond to the $10-11th, 15-16th, 20-21st, 34-35th$ and $43-44th$ pairs of eigenstates
with odd and even parity correspondingly. For the higher excited states the localization is almost
perfect, as can be seen in Figure \ref{Fig12}.

\section{Conclusions}
\label{sec:concl}

\noindent
The few existing investigations on superexponential model systems \cite{Schmelcher2,Schmelcher3,Schmelcher4,Schmelcher5}
have demonstrated that they exhibit a variety of peculiar phenomena and properties. Motivated by this and 
in particular by the importance of the geometry of trapping potentials in a diversity of research problems,
encompassing optical tweezers for soft matter manipulation and processing of atomic degenerate quantum gases,
we have performed in this work a first step towards the exploration of the spectral properties of 
confining superexponential potentials.

\noindent
The original self-interacting superexponential oscillator (SSO) has been shown in ref.\cite{Schmelcher5}
to exhibit a crossover of its eigenvalue spectrum from a scaling behaviour below its transition
point to irregular oscillations above the transition energy. Building on that we started here by
analyzing the spectrum of the SSO with varying amplitude. This way we have illuminated the emergence
of the scaling behaviour of the eigenvalues with increasing amplitude. In a second step we modified
the SSO in several ways in order to demonstrate the variations of the spectrum. For the symmetrized SSO
a smooth approach of the scaled spacing $\frac{\Delta E_{n+1}}{\Delta E_n}$ with increasing degree of
excitation is observed towards the value one, either from above or below one, depending on its amplitude. 
For low excitations a crossover of the scaled spacing from increasing to decreasing behaviour
and from values below one to above one is observed. Augmented by a power law in the exponent we have
analyzed the so-called power law SSO: here the scaled spacing shows an alternating increasing
and decreasing behaviour between neighboring energy spacings of the energy levels. The mean behaviour
is similar to the one obtained for the symmetrized SSO. 

\noindent
In the second part of our investigations we have explored potentials with a spatially oscillating 
power (OPP) in combination with a constant part. These potentials exhibit a central well and an 
infinite series of left and right located side wells whose depths increases with increasing distance
from the central well. We have analyzed the eigenenergy spacing dynamics and revealed several
interesting properties depending on the constant part of the power and the phase of the oscillating
power. The sublinear, linear as well as quadratic and quartic case have been considered. Depending
on the phase these potentials show a reflection symmetry or not. In the presence of the reflection
symmetry a series of energetically near degenerate doublets of eigenstates emerges with an increasingly
smaller splitting with increasing degree of excitation. While the remainder of the eigenstates are
delocalized these doublet states are localized in a left right pair of outer wells. Breaking the
reflection symmetry separates the states of this doublet in energy, i.e. removes the near degeneracy,
and introduces an exclusive left or right localization. 

\noindent 
Based on the above diverse spectral properties of superexponential potentials, there is several
possible future directions of investigation and potential applications. Preparing  as an initial state
a superposition state of the parity symmetric partner eigenstates of a doublet the resulting quantum
dynamics would be an oscillation between right and left localized wave packets, i.e. a remote transfer
or delivery of particles, or, in other words, Rabi oscillations between remotely placed potential
wells. One might then conjecture that superpositioning more than just one doublet,
leads to a 'collective' type of Rabi
oscillations between several occupied outer wells involving different frequencies,
which is an intriguing feature of the OPP. Another relevant question is how interactions would
modify or enrich this picture - meaning both the spectral structure as well as the resulting
quantum dynamics. Related to this general direction of investigation would be the question
of the emergence of entanglement in the OPP. 

\noindent 
Changing the phase of the OPP in time may yield a delivery of particles, even particle by particle,
or atom by atom, to the central well. The OPP could therefore potentially be used as a rechargeable
tweezer where the 'focus' atom to be processed is in the central well and, once this atom is gone,
the central well can be reloaded by changing the phase of the OPP and transfering atoms from the side wells 
to the central well. A very promising physical platform to realize this would be ultracold atoms
which can be prepared, processed and detected on a single atom level \cite{Ott}. The OPP might be created
by one of the extremely powerful techniques of light beam engineering. Traps with almost arbitrary
geometry can be designed nowadays \cite{Grimm} including optical tweezer arrangements \cite{Kuhn}
using spatial light modulators or so-called painted dynamic potentials which can create by a rapidly
moving laser beam time-averaged optical dipole potentials of arbitrary shape \cite{Boshier}.

\section{Acknowledgments}
The author acknowledges K. Keiler for a careful reading of the manuscript.


\begin{thebibliography}{99}
\bibitem{Smith} Low temperatures and cold molecules, Imperial College Press, edited by Ian W M Smith (2008)
\bibitem{Krems} Cold molecules: theory, experiment and applications, CRC Press, edited by R.V. Krems, W.C. Stwalley
and B. Friedrich (2009).
\bibitem{Ho} Handbook of Photonics for Biomedical Engineering, Springer Reference, edited by
A.H.P. Ho, D. Kim, M.G. Somekh (2017).
\bibitem{Pethick} C.J. Pethick and H. Smith, Bose-Einstein condensation in dilute gases, 2nd edition, Cambridge
University Press (2008).
\bibitem{Metcalf} H.J. Metcalf and P. van der Straten, Laser cooling and trapping, Springer (2013).
\bibitem{Grimm} R. Grimm, M. Weidem\"uller and Y.B. Ovchinnikov, Adv.At.Mol.Opt.Phys. 42, 95 (2000).
\bibitem{Chin} C. Chin, R. Grimm, P. Julienne, and E. Tiesinga, Rev.Mod.Phys. 82, 1225 (2010).
\bibitem{Frantzeskakis} D.J. Frantzeskakis, J.Phys.A 43, 213001 (2010).
\bibitem{Moritz} M. Bohlen, L. Sobirey, N. Luick, H. Biss, T. Enss, T. Lompe, and H. Moritz,
Phys.Rev.Lett. 124, 240403 (2020).
\bibitem{Schmelcher1} S. Z\"ollner, H.D. Meyer and P. Schmelcher, Phys.Rev.Lett. 100, 040401 (2008).
\bibitem{Oberthaler} M. Albiez, R. Gati, J. Fölling, S. Hunsmann, M. Cristiani, and M.K. Oberthaler,
Phys.Rev.Lett. 95, 010402 (2005).
\bibitem{Bloch} I. Bloch, J. Dalibard, W. Zwerger, Rev.Mod.Phys. 80, 885 (2008).
\bibitem{Greiner} M. Greiner, O. Mandel, T. Esslinger, T.W. H\"ansch and I. Bloch, Nature 415, 39 (2002).
\bibitem{Schmelcher2} P. Schmelcher, Commun.Nonl.Sci.Num.Sim. 95, 105599 (2021).
\bibitem{Schmelcher3} P. Schmelcher, acc.f.publ. Comm.Nonl.Sci.Num.Sim., arXiv:2011.13672.
\bibitem{Schmelcher4} P. Schmelcher, J.Phys. A 53, 075701 (2020).
\bibitem{Schmelcher5} P. Schmelcher, J.Phys. A 53, 305301 (2020).
\bibitem{Groenenboom} G.C. Groenenboom and H.M. Buck, J.Chem.Phys. 92, 4374 (1990).
\bibitem{Ott} H. Ott, Rep.Prog.Phys. 79, 054401 (2016).
\bibitem{Kuhn} C.Muldoon, L. Brandt, J. Dong, D. Stuart, E. Brainis, M. Himsworth, and A. Kuhn,
New J. Phys. 14, 073051 (2012).
\bibitem{Boshier} K. Henderson, C. Ryu, C. MacCormick and M.G. Boshier, New.J.Phys. 11, 043030 (2009).
\end{thebibliography}
\end{document}